\documentclass[namedreferences]{SolarPhysics}
\usepackage[optionalrh]{spr-sola-addons} 
\usepackage{graphicx}        
\usepackage{color}           
\usepackage{url}             

\newcommand{\doiurl}[1] {\url {http://dx.doi.org/ #1}}
\newcommand{\adsurl}[1] {\url {http://adsabs.harvard.edu/abs/ #1}}

\newcommand{\etal}{{\it et al.}}

\newcommand{\aap}{    {\it Astron. Astrophys.}}

\newcommand{\apj}{    {\it Astrophys. J.}}
\newcommand{\apjl}{    {\it Astrophys. J. Lett.}}

\newcommand{\nat}{    {\it Nature}}

\begin{document}

\begin{article}

\begin{opening}

\title{Variability of Solar Five-Minute Oscillations \\
in the Corona as Observed by the \textit {Extreme \\
Ultraviolet Spectrophotometer} (ESP)\\
on the \textit {Solar Dynamics Observatory Extreme
Ultraviolet Variability Experiment} (SDO/EVE)} 

\author{L.~\surname{Didkovsky}$^{1}$\sep
        A.~\surname{Kosovichev}$^{2}$\sep
        D.~\surname{Judge}$^{1}$\sep
        S.~\surname{Wieman}$^{1}$\sep
        T.~\surname{Woods}$^{3}$
       }
\runningauthor{L.~Didkovsky \etal} \runningtitle{Variability of Five-Minute Oscillations in the Corona }

   \institute{
                $^{1}$ Space Sciences Center, University of Southern California, Los Angeles, USA \\
                     email: \url{leonid@usc.edu}
                     email: \url{judge@usc.edu}  email: \url{wieman@usc.edu} \\
                $^{2}$ Hansen Experimental Physics Laboratory, Stanford University, USA \\
                     email: \url{akosovichev@solar.stanford.edu}\\
                $^{3}$ Laboratory for Atmospheric and Space Physics, University of Colorado at Boulder, USA \\
                     email: \url{tom.woods@lasp.colorado.edu}\\
             }

\begin{abstract}
Solar five-minute oscillations have been detected in the power spectra of two six-day time intervals from soft X-ray measurements of the Sun observed as a star
using  the \textit {Extreme Ultraviolet Spectrophotometer} (ESP) onboard the \textit {Solar Dynamics Observatory} (SDO) \textit {Extreme Ultraviolet Variability Experiment} (EVE).
The frequencies of the largest amplitude peaks were found matching within $3.7~\mu$Hz the known low-degree ($\ell$ = 0\,--\,3) modes
of global acoustic oscillations, and can be explained by a leakage of the global modes into the corona. Due to strong variability of the solar atmosphere between the photosphere and the corona the frequencies and amplitudes of the coronal oscillations are likely to vary with time. We investigate the variations in the power spectra
for individual days and their association with changes of solar activity, {\it e.g.} with the mean level of the EUV irradiance, and its short-term variations due to
evolving active regions. Our analysis of samples of one-day oscillation power spectra for a 49-day period of low and intermediate solar activity showed little correlation with the mean EUV irradiance and the short-term variability of the irradiance.
We suggest that some other changes in the solar atmosphere, {\it e.g.} magnetic fields and/or inter-network configuration may affect the mode leakage to the corona.

\end{abstract}
\end{opening}

\section{Introduction}
     \label{S-Introduction}

The frequencies of individual resonant acoustic modes \cite{Claverie1979} that are excited by turbulent convection
are stable within about $0.4~\mu$Hz on the time scale of the solar cycle as they correspond to intrinsic phase relations of resonant waves
in the solar interior. Five-minute oscillations with frequencies
centered at about 3.3~mHz are trapped below the solar photosphere. However, a number of observations in photospheric and chromospheric lines, in the UV pass-bands, and in the coronal Fe\,{xvi} line (33.5 nm) demonstrate some leakage of these oscillations into upper layers of the solar atmosphere, {\it e.g.} \inlinecite{Judge01}, \inlinecite{O'Shea02}, \inlinecite{McIntosh03}, \inlinecite{Muglach03}, \inlinecite{DePontieu2004}. This leakage may be explained by the increased amplitude of the oscillations due to the rapid density decrease, {\it e.g.} \inlinecite{Gough1993}, and/or by interaction with the network magnetic elements, which can channel the photospheric acoustic power to higher atmospheric layers at frequencies below the cutoff \cite{Vecchio2007}. For a discussion of these observations see \inlinecite{Didkovsky2011}.

\inlinecite{Erdelyi2007} investigated the acoustic response to a single point-source driver. \inlinecite{Malins2007} used a numerical simulation to show that widely horizontally coherent velocity signals from \textit{p}-modes may cause cavity modes in the chromosphere, and surface waves in the transition region, and that fine structures are generated extending from a dynamic transition region into the lower corona, even in the absence of a magnetic field.

A detection of the response of the corona to the observed photospheric low-degree ($\ell$ = 0\,--\,3) \textit{p}-modes was reported by \inlinecite{Didkovsky2011}. The authors studied the oscillation power spectra of two six-day long time-series using the soft X-ray band-pass from the \textit {Extreme Ultraviolet Spectrophotometer} (ESP) \cite{Didkovsky2012} onboard the \textit {Solar Dynamics Observatory} (SDO) \textit {Extreme Ultraviolet Variability Experiment} (EVE: \opencite{Woods2012}). The largest amplitude peaks in the five-minute spectral region were compared with the low-degree photospheric \textit{p}-modes observed in Doppler velocity from the \textit {Birmingham Solar Oscillation Network} (BiSON: \opencite{Chaplin1998}) and in visible light intensity (red channel) from the SOHO/VIRGO instrument \cite{Andersen1991, Frohlich1997}.
This comparison showed that the frequencies of the coronal oscillations may deviate from the frequencies determined from the photospheric observations. This can be
explained by significant influence of the non-uniform distribution of the irradiance sources and variability of the upper atmosphere \cite{Didkovsky2011}. The mean standard deviation of the coronal frequencies from the photospheric \textit{p}-mode frequencies was $\approx 3.7~\mu$Hz in the frequency range of 2.4 to 3.6~mHz, which is about two times larger than the uncertainty of the peaks in the power spectrum determined from the six-day time-series. This deviation was also confirmed by comparing the power spectra for two consecutive six-day time series with a spectrum for the combined 12-day period. The power spectrum for a single six-day time-series showed more significant peaks with better correspondence to the photospheric \textit{p}-mode spectrum than the combined 12-day spectrum. However, it was not clear whether the observed coronal oscillations were transmitted but distorted photospheric \textit{p}-modes, or if these oscillations were excited in the corona by localized impulsive perturbation related to solar activity processes, ({\it e.g.} modeling of \inlinecite{Goode1992}, \inlinecite{Andreev1995}, \inlinecite{Andreev1998}, \inlinecite{Bryson2005}. In the follow-up work here we investigate whether the oscillations were excited in the corona by impulsive sources of solar activity ({\it e.g.} by flares) by studying the observational data during higher solar activity. If they are caused by solar activity, then this study could reveal a correlation between the appearance of the coronal five-minute oscillations and such activity. In contrast to that, if the observed coronal oscillations \cite{Didkovsky2011} were related to the transmission of photospheric \textit{p}-modes through the upper atmosphere, the use of observations made during the higher solar activity may reveal that solar-irradiance variability and significant increases of the soft X-ray irradiance during solar flares add some solar ``noise'' to the data time-series and low-amplitude oscillation peaks in the power spectra may be masked by these noise peaks.

In this work we study the variability of coronal five-minute oscillations by analyzing 49 one-day power spectra for various solar activity observing conditions that range from the lowest solar activity observed during the SDO mission in the middle of May 2010 to intermediate solar activity levels in 2011. As in
\inlinecite{Didkovsky2011} we use
soft X-ray observations without spatial resolution in the zeroth-order channel of SDO/EVE/ESP.

\section{SDO/EVE/ESP Channels}
\label{S-ESP}

EVE \cite{Woods2012} is one of three instrument suites on SDO. It provides solar EUV-irradiance measurements that are unprecedented in terms of spectral resolution, temporal cadence, accuracy, and precision. Furthermore, the EVE program will incorporate physics-based models of solar EUV irradiance to advance the understanding of solar dynamics based on short- and long-term activity of solar magnetic features. ESP \cite{Didkovsky2012} is one of five channels in the EVE suite. It is an advanced version of the SOHO/CELIAS/SEM \cite{Hovestadt95, Judge98}. ESP is designed to measure solar EUV irradiance in four first-order bands of the diffraction grating centered around 19~nm, 25~nm, 30~nm, and 36~nm, and in a soft X-ray band from 0.1 to 7.0~nm (the energy range
is 0.18 to 12.4~keV) in the zeroth-order of the grating. Each band's detector system converts the photo-current into a count-rate (frequency). The count-rates are integrated over 0.25~seconds increments and transmitted to the EVE Science and Operations Center for data processing. An algorithm for converting the measured count rates into solar irradiance and the ESP calibration parameters are described by \citeauthor{Didkovsky07}, (\citeyear{Didkovsky07}, \citeyear{Didkovsky2012}).

\section{Observations} 
      \label{S-Data}
 Our analysis of the ESP measurements was based on datasets for a long series of observations covering a much wider range of solar activity compared to the two six-day time intervals analyzed by \inlinecite{Didkovsky2011}. Due to the high sensitivity of the ESP zeroth-order soft X-ray signal to solar activity and because of significant contamination of the power spectra in the five-minute band by the impulsive increases in irradiance during solar flares, we used data for the periods of small-to-intermediate solar activity, without strong solar flares. Based on these conditions, five data intervals were chosen (Table 1, Figure 1).
  \begin{figure}[ht]
   \begin{center}
   \begin{tabular}{c}
   \includegraphics[height=9.0 cm]{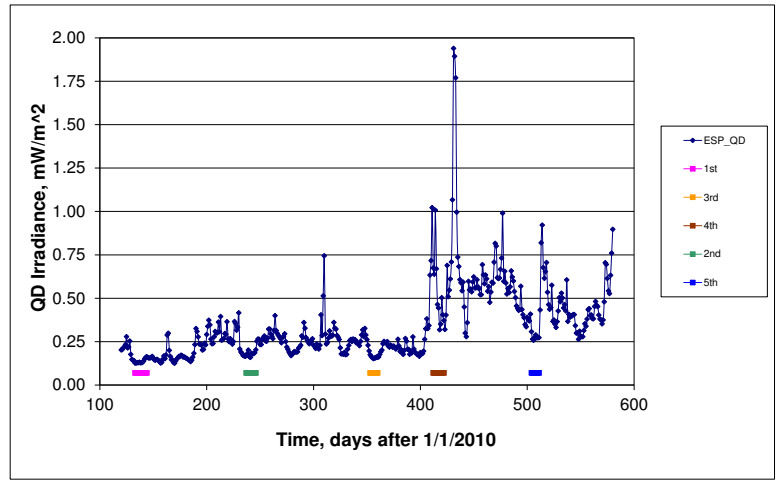}
   \end{tabular}
   \end{center}
   \caption[Figure 1]
   { \label{fig:Figure 1}
ESP zeroth-order (0.1 to 7.0 nm) quad-diode (QD) soft X-ray irradiance measured for the first 461 days of the SDO mission, 245 days in 2010 and 216 days in 2011. The time intervals chosen are shown as wide horizontal bars below the irradiance curve. The mean number of analyzed days for each time interval was about ten, with the total number of days equal to 49. Note, the irradiance curve represents daily mean values while the actual cadence is 0.25~seconds.}
   \end{figure}
 Figure 1 shows the five data intervals used for this analysis (thick horizontal bars) on the background of variations of soft X-ray solar irradiance related to the increased phase of solar activity cycle. Intervals one to three were chosen to represent the lowest periods of solar irradiance observed by SDO in 2010, 0.137, 0.181, and 0.160 mW\,m$^{-2}$, respectively. The fourth time interval was chosen between two periods of relatively high solar activity (Figure 1) with a mean solar irradiance of 0.531 mW\,m$^{-2}$. The fifth time interval represents a return to relatively low solar activity with a mean irradiance of 0.276 mW\,m$^{-2}$. Thus, these five time intervals cover a wide range of solar conditions for the periods of decreased solar activity. To establish a point of reference with our previous analysis, the first time interval matches the two six-day intervals analyzed previously \cite{Didkovsky2011}. Three of the time intervals (the second, third and fifth) correspond to the lower-irradiance ``spots'' on the irradiance curve (Figure 1) with decreased solar activity, while the fourth time interval includes some C and M-class solar flares to have some stronger disturbances of the solar atmosphere. Table 1 shows some details of the five-interval database.
\begin{table}
\caption{Some details of the analyzed datasets. Columns (left to right) are time intervals,
range of days within each time interval, days with GOES class B or larger solar flares, and solar flare class. }
\begin{tabular}{cccc} 
\hline
  Time & Range & Days with  & GOES Solar   \\
  Interval & of Days & Solar Flares & Flare \\
           &  & $\geq B1$  & Class \\
           &         &     &  \\
\hline
1  &  13\,--\,24 May 2010& 13 May 2010 & B2.8  \\
  &  & 22 May 2010 & B1.3   \\
  &  & 23 May 2010  & B1.4, B1.3, B1.1   \\
  &  &24 May 2010  & B1.1   \\
2  &  25 Aug\,--\,3 Sep 2010 & 27 Aug 2010  & B2.2, B1.6, B2.3,   \\
   &              &                 & B4.7, B2.2, B1.9,    \\
   &              &                 & B1.3, B1.1, B2.2,   \\
   &              &                 & B1.9, B1.3, B1.1   \\
  &  &28 Aug 2010 & B1.0   \\
  &  &29 Aug 2010  & B1.7, B1.4   \\
  &  &30 Aug 2010  & B1.4, B1.8  \\
  &  &3 Sep 2010  & B2.8, B1.2   \\
3  &  19\,--\,26 Dec 2010& 19 Dec 2010  & B1.1  \\
4  & 16\,--\,26 Feb 2011& 16 Feb 2011  &C2.0, M1.0, C2.2,   \\
   &            &                  & C5.9, C2.2, M1.1, \\
   &            &                  &  C9.9, C3.2, C1.0,\\
      &            &                &  M1.6, C7.7, C1.3,\\
         &                     &         & C1.1, C4.2, C2.8 \\
  &  &17 Feb 2011  & C6.1, C2.3, C1.5  \\
  &   &                &  C1.0 \\
5 & 19\,--\,26 May 2011 & 19 May 2011  &     \\
 &  & 20 May 2011  & B2.0, B3.5, B3.0,\\
  &                        &         & B2.2 \\
  &  &21 May 2011  & B1.6, B1.1, B2.7   \\
  &  &                & B1.5 \\
  &  & 22 May 2011  & B1.3, B1.2  \\
  &  & 23 May 2011  & B1.9  \\
  &  &24 May 2011  & B2.0   \\
  &  & 25 May 2011  & B1.9, B4.1, C1.1,   \\
  &  &                  & C1.4, B3.0 \\
  &  &26 May 2011  & B1.3, B4.4, B1.9   \\
\hline
\end{tabular}
 \vspace{-0.03\textwidth}
\end{table}
The lowest since the SDO mission daily mean irradiance of 0.126 mW\,m$^{-2}$ was detected for 13 May 2010 (the first day of the first time interval)
and the lowest daily STD of 3.31 $10^{-3}$ mW\,m$^{-2}$ for 15 May 2010. The largest flare-related mean irradiance value and STD were detected for 2011047,
and were 0.674 and 0.288 mW\,m$^{-2}$, respectively.

 \section{Data Reduction}
 \label{S-Reduction}
 The data reduction for this work was based on the use of the ESP zeroth-order time series with original (Level 0D) effective count rate (counts\,s$^{-1}$) corrected for energetic-particle events and temperature changes of dark counts \cite{Didkovsky2012}. The data were interpolated to eliminate short gaps (about two minutes total) that occur when the filters in the ESP filter-wheel and observing modes change during routine daily calibration.
 The first step was to calculate a power spectrum for each of the 49 days analyzed. Then, a power-law curve for each spectrum was determined in a manner similar to that described
 by \inlinecite{Didkovsky2011} for the best fit of the spectrum in the range of frequencies from 2.0~mHz to 10.0~mHz, which includes our range of interest between
 2.4~mHz and 4.0~mHz.
\begin{equation}
\textit{I1}_{i}(f)=A_{i} \times f^{-n_{i}}
\end{equation}
where $\textit{I1}$ is an array for the power-law spectral density, $i$ is the day number, $A_{i}$ is a constant, $f$ is frequency, and $n_{i}$ is the power-law index.
The third step was to calculate a running mean [$RM$] curve which may represent the local, {\it e.g.} in the five-minute band, power increase in the power spectrum.
Due to the use of one-day power spectra with relatively low ($11.6~\mu$Hz) frequency resolution and, thus, a low confidence in the frequencies of individual peaks,
the running mean window of integration was chosen as 13~seconds to reduce the influence of individual peaks in the power spectrum but preserve the whole power increase
within the range of the power-law curve.
\begin{equation}
\textit{I2}_{i}(f)=RM_{i}
\end{equation}
where $\textit{I2}$ is an array which represents the running-mean function. The running mean function is the standard IDL procedure, \textit{i.e.} median.
The final step was to calculate the mean sum $S_{i}$ of the ratios between the $I2_{i}$ and $I1_{i}$ over the frequency range of five-minute oscillations ($f_{1}=2.4$~mHz to $f_{2}=4.0$~mHz).
\begin{equation}
S_{i}=\frac{\sum_{f_{1}}^{f_{2}} \frac{\textit{I2}_{f}}{\textit{I1}_{f}}}{ m}
\end{equation}
where $m$ is the number of frequency bins within the frequency range.
If the power spectrum showed an increase in the five-minute band, then $S_{i}$ is greater than unity.
As an example of this data reduction algorithm, Figure 2 shows a power spectrum for 13 May 2010 with $\textit{I1}$ as a dashed (blue) line and $\textit{I2}$ as a red line.
  \begin{figure}[ht]
   \begin{center}
   \begin{tabular}{c}
   \includegraphics[height=10.0 cm]{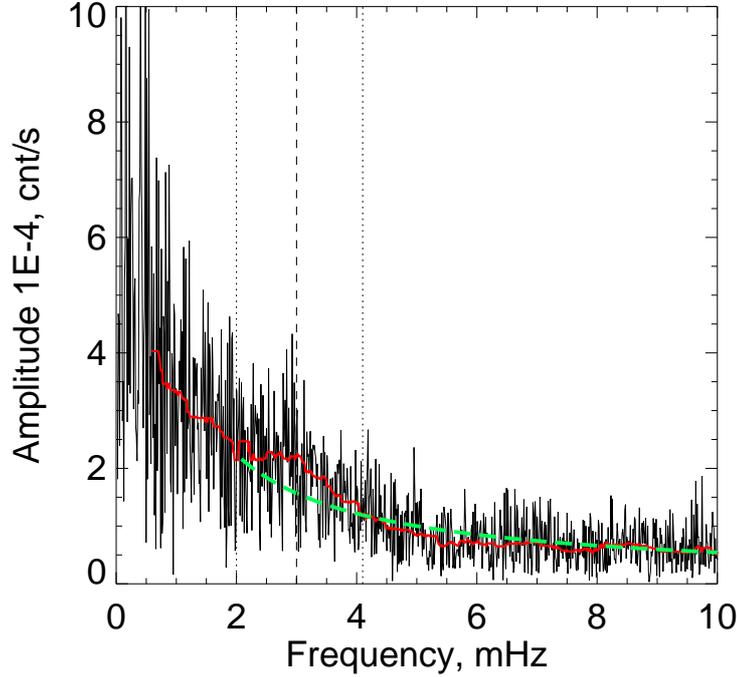}
   \end{tabular}
   \end{center}
   \caption[Figure 2]
   { \label{fig:Figure 2}
ESP zeroth-order (0.1 to 7.0 nm) power spectrum for 13 May 2010 (2010 DOY 133). Dashed (green) line is a power-law curve ($\textit{I1}=0.18 f^{-0.88}$),
 red line ($\textit{I2}$) is a running-mean curve. Two dotted vertical lines represent the edges of the power increase and the dashed vertical line shows the frequency of the maximum of the increase (see Table 4). The value of $S$ for this example is 1.23.}
   \end{figure}
 In addition to the power increase in the five-minute region, some peaks with frequencies above the cut-off frequency of 5.5~mHz demonstrate increased amplitudes without the clear indication of increased power represented by the red line (Figure 2). For some other days the power increase in the frequency region from 6 to 10~mHz is clearly observed and this increase may be similar to that found by \inlinecite{Gurman82} in the frequency range from 5.8 to 7.8 mHz using SMM observations. An example of such increases in the five-minute region and above the cut-off frequency is shown in Figure 3.
  \begin{figure}[ht]
   \begin{center}
   \begin{tabular}{c}
   \includegraphics[height=10.0 cm]{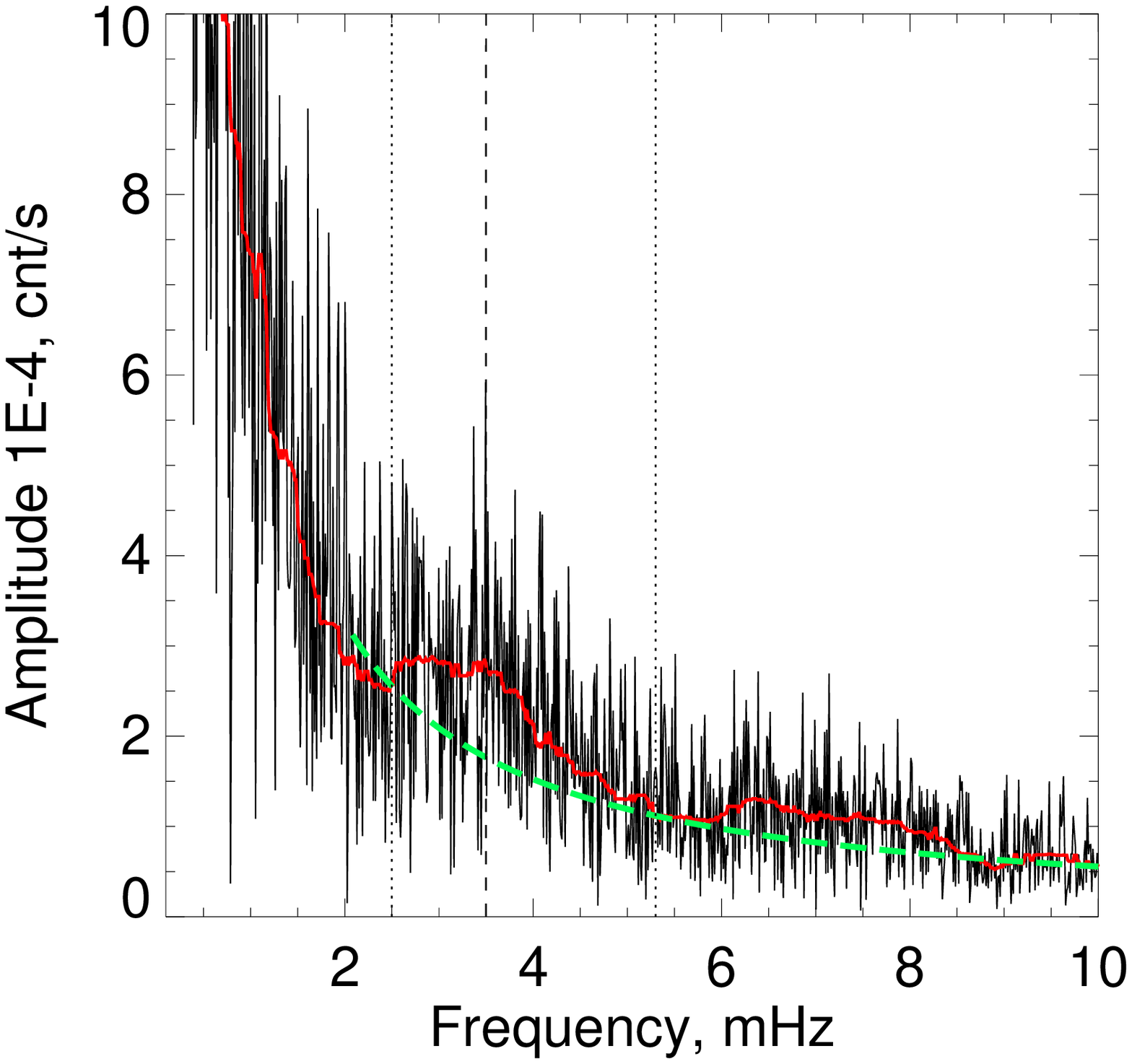}
   \end{tabular}
   \end{center}
   \caption[Figure 3]
   { \label{fig:Figure 3}
ESP zeroth-order (0.1 to 7.0 nm) power spectrum for the 23 May 2010 (DOY 142). The dashed line is a power-law curve ($\textit{I1}=7.0 10^{-4} \times f^{-1.1}$),
the red line ($\textit{I2}$) is a running-mean curve. Two dotted vertical lines represent the edges of the power increase and the dashed vertical line shows the frequency corresponding to the maximum of the increase (see Table 4). The $S$-ratio for this example is 1.37. Power increases above the cut-off frequency, at about 6 -- 8~mHz are more visible than in Figure 2. Note, for the convenience of comparison with Figure 2, the vertical scale is the same as in Figure 2.}
   \end{figure}

\section{Results} 
      \label{S-Results}
The results of the calculation of the five-minute $S$-ratios for the one-day power spectra are shown in Table 2 along with the changes of spectral irradiance, fluctuations of this irradiance (STD), and the maximum amplitude of the filter curve $\textit{I2}$ (Equation (2)) in the five-minute range.
\begin{table}
\caption{A comparison of the $S$-ratios (Equation (3)) in the five-minute band with the daily mean soft X-ray irradiance, the standard deviation of this irradiance (STD), and the filter $\textit{I2}$ maximum amplitude (Equation (2)).}
\begin{tabular}{cccccc} 
\hline
  Day, & Daily mean,  & Standard  & $S$ & $\textit{I2}$  \\
   YYYY\,DOY & Irradiance & Deviation &  &\\
\hline
2010\,\,133 & 1.26 & 7.54 &  1.23 & 2.3 \\
2010\,\,134 & 1.26 & 3.83 &  1.07 & 0.9 \\
2010\,\,135 & 1.29 & 3.31 &  1.14 & 1.4 \\
2010\,\,136 & 1.28 & 3.40 &  0.87 & \\
2010\,\,137 & 1.32 & 3.99 &  1.12 & 1.1 \\
2010\,\,138 & 1.28 & 3.47 &  1.06 & 1.1 \\
2010\,\,139 & 1.29 & 3.67 &  1.05 & 1.5 \\
2010\,\,140 & 1.30 & 3.90 &  1.07 & 0.9 \\
2010\,\,141 & 1.39 & 7.31 &  1.00 & 1.5 \\
2010\,\,142 & 1.52 & 9.58 &  1.37 & 2.9 \\
2010\,\,143 & 1.60 & 18.4 &  1.03 & 2.5 \\
2010\,\,144 & 1.63 & 11.7 &  1.01 & 2.5 \\
\hline
 2010\,\,237 & 1.69 & 19.50 & 0.99  & \\
 2010\,\,238 & 1.81 & 18.40 & 0.98  & \\
 2010\,\,239 & 2.01 & 27.2 &   N/A &  \\
 2010\,\,240 & 1.63 & 8.09 &  1.11 & 2.4 \\
 2010\,\,241 & 1.59 & 6.68 &  1.05 & 2.4 \\
 2010\,\,242 & 1.83 & 11.5 &  1.07 & 1.9 \\
 2010\,\,243 & 1.80 & 6.96 &  0.98 &     \\
 2010\,\,244 & 1.81 & 5.43 &  1.03 & 1.7 \\
 2010\,\,245 & 1.86 & 8.87 &  1.08 & 2.4 \\
 2010\,\,246 & 2.03 & 13.6 &  1.02 & 3.1 \\
\hline
 2010\,\,353 & 1.74 & 6.14 & 0.98  &\\
 2010\,\,354 & 1.64 & 6.24 & 0.95  &\\
 2010\,\,355 & 1.58 & 3.98 & 0.96  &\\
 2010\,\,356 & 1.52 & 3.41 & 0.78  &\\
 2010\,\,357 & 1.55 & 4.47 & 0.95  &\\
 2010\,\,358 & 1.56 & 4.91 & 0.88  &\\
 2010\,\,359 & 1.62 & 5.09 & N/A  & \\
 2010\,\,360 & 1.61 & 4.68 & 0.90  & \\
\hline
 2011\,\,047 & 6.74 & 288.0 & 1.24  & 225 \\
 2011\,\,048 & 6.38 & 147.0 & 0.95  & \\
 2011\,\,049 & 1.01 & 504.0 & N/A  & \\
 2011\,\,050 & 6.70 & 178.0 & 1.16 & 108 \\
 2011\,\,051 & 4.64 & 75.9 & 0.97  & \\
 2011\,\,052 & 4.44 & 112.0 & N/A  & \\
 2011\,\,053 & 3.18 & 25.1 & 0.96  & \\
 2011\,\,054 & 3.50 & 42.2 & 0.93  & \\
 2011\,\,055 & 5.03 & 505.0 & N/A  & \\
 2011\,\,056 & 4.05 & 36.9 & N/A  & \\
 2011\,\,057 & 3.72 & 60.6 & 1.08  & 28.4 \\
\hline
 2011\,\,139 & 3.08 & 30.8 &  N/A  & \\
 2011\,\,140 &  2.65 & 9.3 & 0.95  & \\
 2011\,\,141 & 2.58 & 6.37 & N/A   & \\
 2011\,\,142 & 2.72 & 9.66 & N/A   & \\
 2011\,\,143 & 2.89 & 10.6 & N/A   & \\
 2011\,\,144 & 2.72 & 12.8 & 1.02  & 3.3 \\
 2011\,\,145 & 2.73 & 32.1 & N/A   & \\
 2011\,\,146 & 2.73 & 20.1 & N/A   & \\
\hline
\end{tabular}
 \vspace{-0.03\textwidth}
\end{table}
Columns (one to five) are days of observations in the (YYYY\,\,DOY) format, Daily mean irradiance $[10^{-1}$ mW\,m$^{-2}]$, Daily Standard deviation $[10^{-3}$ mW\,m$^{-2}]$, the ratio ($S$), and the filter $\textit{I2}$ maximum amplitude in the five-minute range, $[10^{-4}$ counts\,s$^{-1}]$. Each of the five time intervals is separated by horizontal lines.
Note, the $S$-ratio column in Table 2 is marked as N/A if the change of the spectral density in the power spectrum shows large low-frequency variations that could be caused by strong changes of solar irradiance, {\it e.g.} related to a solar flare. The gaps in the $\textit{I2}$ column are either for days in which the power spectra do not show any increase in the five-minute region (the $S$-ratio is $<1$) or for days where the $S$-ratio in the $S$-ratio column is marked as N/A because of strong contamination from solar flares. As Table 2 shows, the largest number of such contaminated spectra corresponds to the fourth and fifth time intervals with significantly higher solar activity (see the STD in Table 2) than during the first three time intervals.

\subsection{How Solar Activity Affects the Oscillations in the Corona } 
      \label{S-Analysis}

To analyze how the oscillations in the five-minute band represented by the $S$-ratios (Table 2) and by the amplitude of the filter curve $\textit{I2}$ (Equation (2)) are related to the changes of the observing conditions, two parameters of these conditions, daily mean irradiance and standard deviation of the irradiance were compared with the changes of the $S$-ratios and maximum amplitude of filter $\textit{I2}$. Table 2 shows that the $S$-ratios larger than unity are detected for the first, second, and fourth time series.
Figures 4\,--\,6 show these time series. Note, the column $\textit{I2}$ in Table 2 for the amplitude of the filtered curve and the open circles in Figures 4\,--\,6 do not show any negative amplitude compared to the power law curve, (Equation (1)) for the days for which the $S$-ratio is less than unity. The filtered curve $\textit{I2}$ (Equation (2)) for such days is below the spectral density $\textit{I1}$ (Equation (1)) and represents nothing but noise. The $S$-ratios (Table 2) are plotted according to a linear scale shown on the right-hand edge of the plots.
  \begin{figure}[ht]
   \begin{center}
   \begin{tabular}{c}
   \includegraphics[height=9.0 cm]{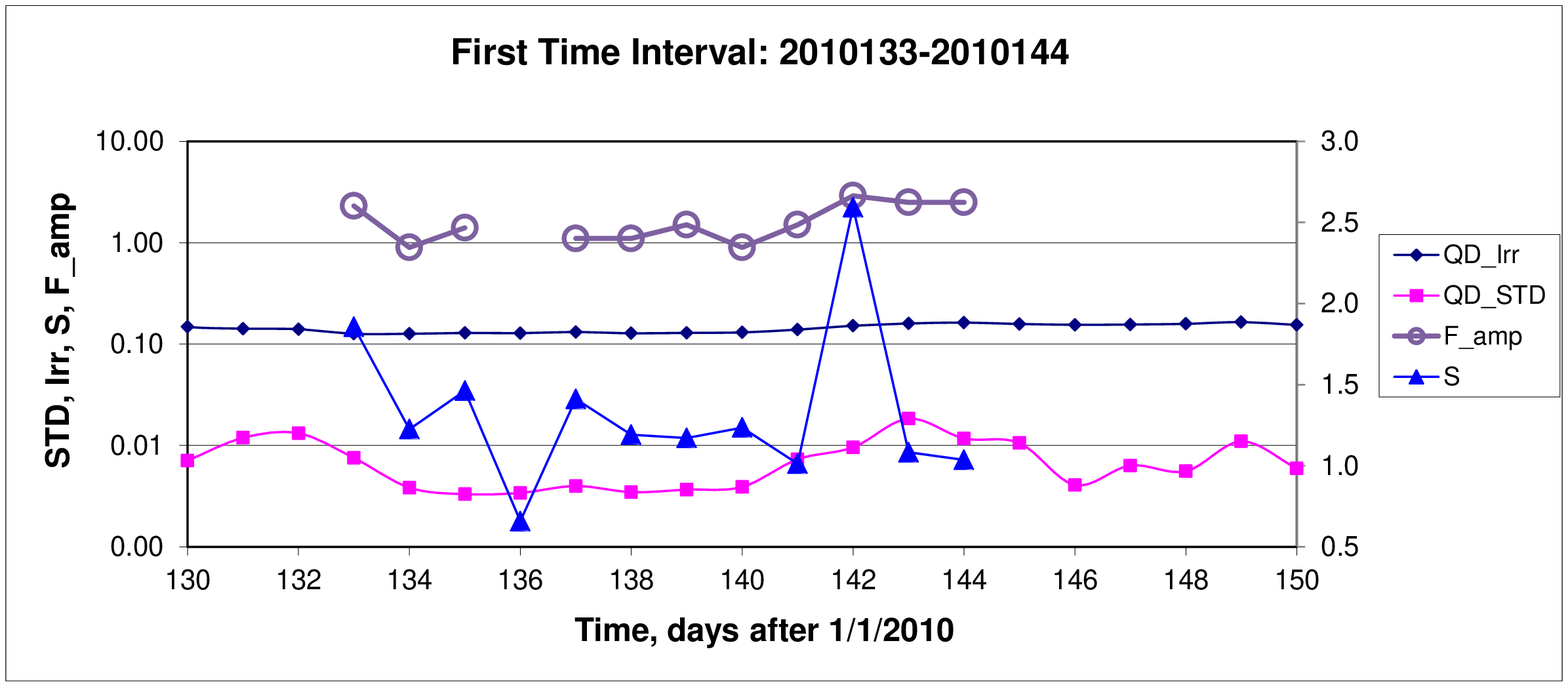}
   \end{tabular}
   \end{center}
   \caption[Figure 4]
   { \label{fig:Figure 4}
Time series for the first time interval used to analyze the correlation between the five-minute oscillation ratio variability (triangles), the changes of the daily mean irradiance (diamonds), and the STD variations of the irradiance (squares). Some positive correlation, {\it e.g.} between days 133\,--\,134 and 141\,--\,142 is changing to  anti-correlation, see {\it e.g.} 140\,--\,141 and 142\,--\,143. The filter $\textit{I2}$ (Equation (2)) maximum amplitudes (open circles) within the five-minute range are shown for the increases of the ratio $S > 1.0$ (right scale). The gap in the maximum amplitude of filter $\textit{I2}$ (open circles) for day 136 in Figure 4 and other gaps in Figures 5 and 6 are related to the ratios with $S < 1.0$ (Table 2, $S$-ratio) for which no power increase and the spectra show noise in the five-minute region.}
   \end{figure}
  \begin{figure}[ht]
   \begin{center}
   \begin{tabular}{c}
   \includegraphics[height=9.0 cm]{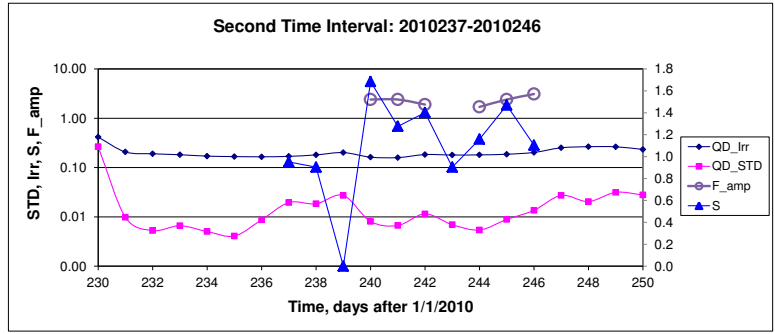}
   \end{tabular}
   \end{center}
   \caption[Figure 5]
   { \label{fig:Figure 5}
   Time series for the second ten-day time interval (2010\,DOY\,237\,--\,246) used to analyze the correlation between the five-minute oscillation ratio variability (triangles), the changes of the daily mean irradiance (diamonds), and the STD variations of the irradiance (squares). Some positive correlation, {\it e.g.} between days 237\,--\,238, 240\,--\,241, and 242\,--\,243 is changing to  anti-correlation, see {\it e.g.} 238\,--\,239, 243\,--\,244, and 245\,--\,246. The filter $\textit{I2}$ (Equation (2)) maximum amplitudes (open circles) within the five-minute range are shown for the increases of the ratio $S > 1.0$ (right scale). The gaps in the maximum amplitude of filter $\textit{I2}$ (open circles) for days 237\,--\,239, and 243 in Figure 5 are related to the ratios with $S < 1.0$ (Table 2, $S$-ratio) for which no power increase and the spectra show noise in the five-minute region.}
   \end{figure}
  \begin{figure}[ht]
   \begin{center}
   \begin{tabular}{c}
   \includegraphics[height=9.0 cm]{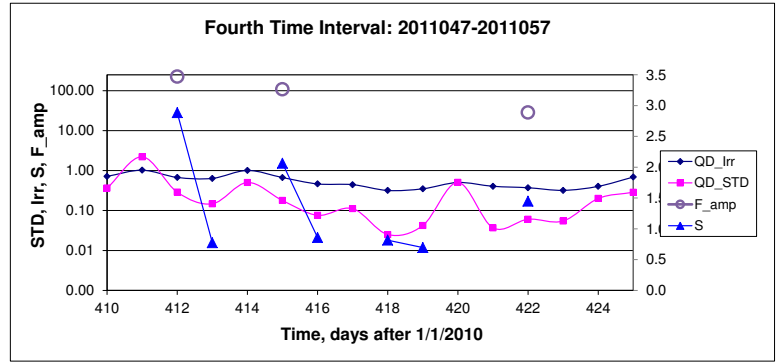}
   \end{tabular}
   \end{center}
   \caption[Figure 6]
   { \label{fig:Figure 6}
   Time series for the fourth 11-day time interval (2011\,DOY\,47\,--\,57) shown as days 412\,--\,422 used to analyze the correlation between the five-minute oscillation ratio variability (triangles), the changes of the daily mean irradiance (diamonds), and the STD variations of the irradiance (squares). Some positive correlation, {\it e.g.} between days 412\,--\,413 and 415\,--\,416 is changing to anti-correlation, see {\it e.g.} 418\,--\,419. The filter $\textit{I2}$ (Equation (2)) maximum amplitudes (open circles) within the five-minute range are shown for the increases of the ratio $S > 1.0$ (right scale). The gaps in the maximum amplitude of filter $\textit{I2}$ (open circles) for days 413\,--\,414, and 416\,--\,421 in Figure 6 are related to the ratios with $S < 1.0$ (Table 2, $S$-ratio) for which no power increase and the spectra show noise in the five-minute region.}
   \end{figure}

Table 3 summarizes data from Table 2 for a more detailed comparison of the observing conditions.
\begin{table}
\caption{A comparison of the $S$-ratios in the five-minute band and their variations with the mean irradiance for each of the five time intervals. }
\begin{tabular}{cccc} 
\hline
  Days, & Mean    & Mean  & Mean    \\
  YYYY\,\,DOY & Irradiance & STD   &   $S$-ratio  \\
           &  10$^{-1}$ [mW\,m$^{-2}$] &  10$^{-3}$[mW\,m$^{-2}$]  & \\
\hline
 2010\,\,133\,--\,144 & 1.37 $\pm$ 1.35 &  6.68 $\pm$ 4.69 & 1.09 $\pm$ 0.12 \\
 2010\,\,237\,--\,246 & 1.81 $\pm$ 1.45 &  12.6 $\pm$ 7.07 & 1.03 $\pm$ 0.05 \\
 2010\,\,353\,--\,360 & 1.60 $\pm$ 0.07 &  4.86 $\pm$ 0.97 & 0.92 $\pm$ 0.07 \\
 2011\,\,047\,--\,057 & 5.31 $\pm$ 2.02 &  180  $\pm$ 178  & 1.04 $\pm$ 0.12 \\
 2011\,\,139\,--\,146 & 2.76 $\pm$ 0.15 &  16.5 $\pm$ 10.1 & 0.99 $\pm$ 0.05 \\

\hline
\end{tabular}
 \vspace{-0.03\textwidth}
\end{table}

The technique used for this analysis is based on a comparison of spectral amplitudes in the five-minute range of the power spectra. This technique is very sensitive to the contamination of the spectra by flare-related increases of the irradiance. These increases affect both the daily mean solar irradiance and the standard deviation [STD] of this irradiance. If the local five-minute increase in the power spectrum is a result of such contamination (flare-related solar noise), one should expect a positive correlation between the $S$-ratio (or maximum amplitude of filter $\textit{I2}$) and the flare-related increases of the irradiance and its STD. Thus, the technical goal of this analysis was to extract and compare such information from the daily spectra for different levels of solar activity.

\subsubsection{Maximum Amplitude of Filter $\textit{I2}$ } 
      \label{S-Amplitude}

Figures 4\,--\,6 show the maximum amplitude of filter $\textit{I2}$ as open circles (see also last column in Table 2). With conditions of minimum solar activity, the oscillation peaks in the power spectra are not masked by noise and we assume that the maxima in the amplitude of filter $\textit{I2}$ for the first and second time intervals show the amplitudes of the five-minute oscillations in the corona related to the photospheric \textit{p}-modes. For such conditions the mean amplitude is similar for the first and second time intervals, 1.69 and $2.16 \times 10^{-4}$ counts\,s$^{-1}]$. The fourth time interval shows significantly higher amplitudes (Figure 6 and Table 2) which indicate the contamination of the spectra by much higher solar flare activity. For low
solar-activity periods the maximum amplitude of filter $\textit{I2}$ is just another representation of the $S$-ratio, {\it e.g.} Figure 4.

\subsubsection{Daily Mean Irradiance } 
      \label{S-Irradiance}

Table 2 and Figures 4\,--\,6 show that the daily mean soft X-ray irradiance is a significant source of the $S$-ratio change. Table 3 shows that the largest mean $S$-ratio of the oscillations in the five-minute band was detected for the first time interval with the lowest daily mean irradiance. Cross-correlation between the changes of the irradiance and the $S$-ratios for the first time interval is low: 0.1. For the second time interval it becomes negative, -0.27, which indicates that an increase of irradiance (1.81 compared to 1.37) leads to a decrease of $S$-ratios for the observed five-minute oscillations in the corona. For the fourth time interval the correlation is positive and high: 0.64. We interpret this fact as a contamination of the power spectra by solar-flare events. The amplitudes of the filter-curve maxima in Figures 4\,--\,6 (open circles) and the last column in Table 2 demonstrate such a flare-related increase. Since the spectral contamination is the result of transfer of the solar flare low-frequency power to the other frequency regions of the spectrum, including the five-minute region, we assume that it leads to a large positive correlation detected for the fourth time interval.

\subsubsection{Standard Deviation } 
      \label{S-STD}

The soft X-ray signal that ESP detects in the zeroth-order channel is a very sensitive probe of solar variability. Assuming that the observed five-minute oscillations in the corona represent a response of the corona to the photospheric acoustic modes, and that the ``transmission'' of the solar atmosphere is a function of various disturbances and inhomogeneities between the photosphere and the corona, we can treat the standard deviation [STD] as an indicator of such solar ``noise'' in the five-minute oscillation signal. However, our results indicate that STD is not a unique parameter to estimate this ``transmission''. For example, the cross-correlation between the $S$-ratio and STD for the first and second time intervals, 0.1 and -0.44, correspondingly, is either low or negative. For the third time interval for which STD was the lowest (Table 3) the $S$-ratios were all $<$ 1.0 (Table 2). This may indicate that, as suggested by a number of authors {\it e.g.} \inlinecite{Judge01}, \inlinecite{O'Shea02}, \inlinecite{McIntosh03}, \inlinecite{Muglach03}, \inlinecite{Vecchio2007}, the connectivity between the photosphere and the corona depends on the configuration of the magnetic fields which may be also a function of the STD. If the lowest STD during the third time interval is related to the decreased magnetic field strength and to a non-effective configuration of the network, this may explain the absence of the power increases in the five-minute range of oscillations. Relatively high negative correlation (-0.44) for the second time interval with about two times larger STD compared to the first time interval allows us to consider that $S$-ratios for the second time interval as well as for the first time interval were not the result of the spectral contamination in the power spectra. This evidence is consistent with another independent confirmation of the response of the corona to photospheric \textit{p}-modes {\it e.g.} based on the spectral analysis of the two six-day time series  \cite{Didkovsky2011}.

\subsubsection{A Shift of Maximum Frequency as a Function of Activity } 
      \label{S-Shift}

The oscillation spectra with the most significant power increases ($S \geqslant 1.1$, see Table 2, $S$) in the five-minute range were analyzed to investigate the correlation between the frequency of the maximum of this increase and solar activity (see Table 2, Standard Deviation). In addition to this correlation, the ranges of the increases were also analyzed. Table 4 and Figure 5 show the results of this analysis.
\begin{table}
\caption{A comparison of the STD, the maximum of the frequency, and the ranges of power increases for parameter $S \geqslant 1.1$. }
\begin{tabular}{ccccc} 
\hline
  Days, & STD of the   & Max  & Range of   & Mean\\
  YYYY\,\,DOY & Irradiance & Frequency   &  Frequency & Frequency  \\
           & [10$^{-6}$ W m$^{-2}$]   &  [mHz]  & [mHz] & [mHz] \\

\hline
 2010\,\,133 & 7.54 &  3.0 & 2.0\,--\,4.1 & 3.05 \\
 2010\,\,135 & 3.31 &  2.4 & 2.1\,--\,4.1 & 3.1  \\
 2010\,\,137 & 3.99 &  3.0 & 2.6\,--\,3.5 & 3.05 \\
 2010\,\,142 & 9.58 &  3.5 & 2.5\,--\,5.3 & 3.9 \\
 2010\,\,240 & 8.09 &  2.7 & 2.1\,--\,4.4 & 3.25 \\
 2011\,\,047 & 288  &  3.6 & 2.4\,--\,5.4 & 3.9 \\

\hline
\end{tabular}
 \vspace{-0.03\textwidth}
\end{table}

  \begin{figure}[ht]
   \begin{center}
   \begin{tabular}{c}
   \includegraphics[height=9.0 cm]{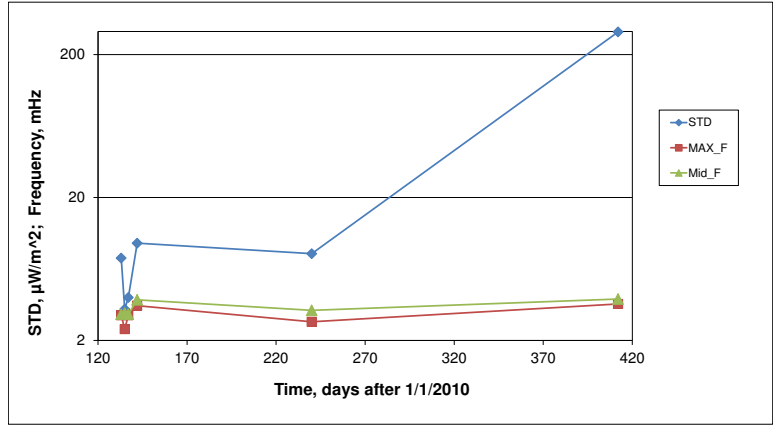}
   \end{tabular}
   \end{center}
   \caption[Figure 7]
   { \label{fig:Figure 7}
A comparison of the changes of solar activity [STD: diamonds] with the changes of the maximum of the frequency increase [squares], and the mean frequency of the range of the increase [triangles]. Note, data for the power spectra are shown with increases of $S \geqslant 1.1$, see Table 2, $S$-ratio. }
   \end{figure}
 Figure 7 shows a positive correlation between the increases of the STD and the shift of the maximum frequency, $R_{1} = 0.62$. The same correlation ($R_{2} = 0.66$) is found between the changes of STD and the mean frequency of the increase. Certainly, the significances of these correlations are small. With the use
 of the \textit{t}-distribution, $t= 1.4$, which is lower than the \textit{t}-distribution number of 1.638 for three degrees of freedom and
 significance 0.1 (one tail). Two days with the largest STD show a significant shift of the right edge of the frequency range toward the cut-off frequency, see 2010 DOY\,142 (5.3 mHz) and 2011 DOY\,47 (5.3 mHz) in Table 4. This shift is consistent with the model proposed by \inlinecite {Bryson2005}.

 \section{Concluding Remarks }
     \label{S-Conclusions}
 Global solar oscillations in the five-minute range were initially detected in the corona using soft X-ray irradiance measurements from SDO/EVE/ESP \cite{Didkovsky2012}. The variability of the five-minute $S$-ratios (Equation (3)) in the power spectra characterizing the strength of oscillations relative to noise was analyzed for five time intervals with low and intermediate solar activity during a 379-day period (2010 DOY 133\,--\,2011 DOY 146) for 49 days.

The results of this analysis show that the best conditions for observing five-minute solar oscillations in the corona are when the solar activity is low. In the first time interval 11 of the 12 days show power increases with an estimated $S$-ratio $\geq 1.0$.

Our analysis shows that power increases in the spectra of coronal oscillations, interpreted as the response of the corona to photospheric \textit{p}-modes due to their channeling \cite{Vecchio2007} or leakage \cite{Erdelyi2007, Malins2007}, are not related to increased daily mean solar irradiance. This is clear from $S$-ratio comparison between the first time interval and the other four time intervals, see Table 3. The larger daily mean irradiances observed for the second through fifth time intervals did not lead to a larger $S$-ratio.

The power increases in the oscillation spectra are not caused by the increases in the mean STD (third column in Table 3) either. This conclusion is based on the analysis of cross-correlations between the $S$-ratio and the mean STD. The correlation is low for all time intervals except the fourth. This result confirms that detected increases of the $S$-ratios in the five-minute range are not created by the spectral contamination in the power spectra from solar flares, and are not the instrumental or data reduction artifacts but represent the leakage of photospheric \textit{p}-modes to the corona.

We interpret a significant positive correlation for the fourth time interval as an artifact and a reflection of much higher solar activity with the STD about 27 times larger than for the first time interval. The maxima of amplitude of filter $\textit{I2}$ for the fourth time interval are significantly larger than the amplitudes for the first and second time intervals and may be a demonstration of contamination of the spectra in the five-minute region by the low-frequency flare signals. This is also clear from helioseismology results which have shown shifts in the frequencies of the global modes related to the solar cycle but not in the amplitudes.

Another conclusion based on these results is that large-scale solar oscillations detected in the corona
are related to the leakage of the photospheric acoustic oscillations rather than to the excitation of these coronal oscillations by solar energetic events.

The high sensitivity of the five-minute oscillations in the corona to the changes of solar irradiance may be used as a diagnostic tool to characterize
the `connectivity' in soft X-ray irradiance dynamics.

\begin{acks}
This work was partially supported by the University of Colorado award 153-5979.
Data courtesy of NASA/SDO and the EVE science team.
\end{acks}


\end{article}

\end{document}